\documentclass[prb,twocolumn]{revtex4}%
\usepackage{amsfonts,verbatim,bm,mathrsfs}
\usepackage{amsmath}
\usepackage{amssymb}
\usepackage{graphicx}
\usepackage{epstopdf}
\usepackage{amsfonts}%
\usepackage[normalem]{ulem}
\begin{document}
\title{Zero bias conductance peak in Majorana wires made of
semiconductor-superconductor hybrid structures}
\author{Chien-Hung Lin$^{1}$}
\author{Jay D. Sau$^{2}$}
\author{S. Das Sarma$^{1}$}
\affiliation{$^{1}$Condensed Matter Theory Center and Joint Quantum Institute, Department
of Physics, University of Maryland, College Park, Maryland 20742-4111, USA.}
\affiliation{$^{2}$Department of Physics, Harvard University, Cambridge, MA 02138, USA.}
\date{\today}

\begin{abstract}
Shell document for REV\TeX{} 4.

\end{abstract}
\begin{abstract}
Motivated by a recent experimental report \cite{1} claiming the likely
observation of the Majorana mode in a semiconductor-superconductor hybrid
structure \cite{2,3,4,5}, we study theoretically the dependence of the zero
bias conductance peak associated with the zero-energy Majorana mode in the
topological superconducting phase as a function of temperature, tunnel barrier
potential, and a magnetic field tilted from the direction of the wire for
realistic wires of finite lengths. We find that higher temperatures and tunnel
barriers as well as a large magnetic field in the direction transverse to the
wire length could very strongly suppress the zero-bias conductance peak as
observed in Ref.[\onlinecite{1}]. We also show that a strong magnetic field
along the wire could eventually lead to the splitting of the zero bias peak
into a doublet with the doublet energy splitting oscillating as a function of
increasing magnetic field. Our results based on the standard theory of
topological superconductivity in a semiconductor hybrid structure in the
presence of proximity-induced superconductivity, spin-orbit coupling, and
Zeeman splitting show that the recently reported experimental data are
generally consistent with the existing theory that led to the predictions for
the existence of the Majorana modes in the semiconductor hybrid structures in
spite of some apparent anomalies in the experimental observations at first
sight. We also make several concrete new predictions for future observations
regarding Majorana splitting in finite wires used in the experiments.

\end{abstract}
\maketitle

\section{\bigskip Background}

The search for solid state Majorana modes, which are localized quasiparticles
with non-Abelian braiding statistics and a direct realization of the Majorana
operator, has created a great deal of recent interest in the whole physics
community\cite{R1}, partly because of the great mystique associated with
Majorana and partly because of the prospects for topological quantum
computation using solid state Majorana quasiparticles. The most experimentally
promising among the many theoretical proposals for solid state Majorana
fermions is the semiconductor-superconductor hybrid structures\cite{2,3,4,5}
where the ordinary s-wave superconducting proximity effect induced in the
semiconductor is modified by the intrinsic spin-orbit coupling in the
semiconductor and an externally applied Zeeman spin splitting. The solution of
the resultant superconducting gap equations, often called the Bogulibov-De
Gennes (BdG) equations, in the presence of both spin-orbit coupling and spin
splitting explicitly shows a quantum phase transition in the nature of the
superconducting phase as a function of the Zeeman splitting (which can be
controlled by an external magnetic field). The lower magnetic field part of
the superconducting phase is an ordinary non-topological s-wave superconductor
with a suppressed gap due to the finite spin splitting which eventually gives
way to a topological (effectively a p-wave) superconducting phase at large
spin splitting when the original proximity-induced s-wave gap is completely
suppressed. The Majorana mode appears naturally as a stable zero-energy
solution inside the gap of this topological superconducting phase. It has been
shown that the existence of the Majorana quasiparticle would give rise to a
conductance peak at zero energy inside the nominal p-wave superconducting gap
of the topological phase. The experimental observation of this predicted zero
energy conductance peak inside the superconducting gap at finite magnetic
field in the semiconductor-superconductor hybrid structure is therefore
believed to be the necessary evidence in support of the solid state Majorana
mode\cite{1}.

It is useful to compare our semiconductor nanowire model for the Majorana modes
with the original 1D Majorana wire model introduced by Kitaev ("the Kitaev model"). 
In the Kitaev model, the  topological $p_x$-wave superconductivity 
%  (and hence of
%the localized Majorana modes at the two ends of the wire)
 is assumed explicitly by making the system a spinless p-wave superconductor.  
% In the Kitaev model, there is
%no topological phase transition and there is no non-topological trivial phase-- the
%Majorana is always there by construction. 
 In fact, Kitaev in his paper \cite{10} goes on to discuss at quite some length the possibility of realizing such
a topological 1D spinless p-wave superconductivity in nature, concluding that the
most suitable systems for realizing such spinless 1D p-wave superconductivity are
quasi-1D organic superconductors involving charge and spin density wave ground
states, which were already proposed as possible Majorana-carrying systems by
Sengupta et al. \cite{9} who in fact also pointed out that the zero bias conductance peak
associated with such Majorana modes would be quantized.  It is a curious historical
coincidence that Kitaev in his original paper explicitly rejected both spin-orbit
coupling and Zeeman splitting as possible physical mechanisms for producing 1D
spinless p-wave superconductivity!  The current activity in 1D semiconductor
Majorana wires started from the work of Sau et al. \cite{2} who showed that in
2D semiconductors with proximity s-wave superconductivity induced by a nearby
ordinary s-wave superconducting metal, the topological superconductivity ("2D chiral
p-wave superconductor") would naturally arise in the presence of spin splitting and
spin-orbit coupling provided that the spin-splitting is large enough to overcome the
trivial s-wave superconductivity.  It was soon realized \cite{3,4,5}
 that this 2D chiral topological superconductivity can easily be modified
to 1D helical topological superconductivity by considering 1D semiconductor
nanowires rather than 2D semiconductor heterostructures with the spin splitting
being introduced by an external magnetic field.  This helical 1D p-wave
superconductor is an effective realization of the 1D Kitaev model, but it is not
identical to the Kitaev model.  For example, the topological superconductivity in
the 1D nanowires exists only above a finite value of the spin splitting with only
trivial non-topological superconductivity existing at weaker magnetic field values. 
Also, the superconducting gap depends explicitly on the spin splitting and the
spin-orbit coupling in the semiconductor nanowires instead of being just effective
theoretical parameters.

\section{\bigskip Introduction}

In a recent presentation \cite{1}, the likely experimental observation of the
theoretically-predicted \cite{2,3,4,5} zero-energy Majorana modes in
semiconductor nanowires, in close proximity to an ordinary (i.e. s-wave)
superconductor and in the presence of an external magnetic field applied along
the wire, has been reported. This experimental observation of the predicted
zero bias peak and the associated implication that this may finally be the
real-life evidence for the existence of the elusive (and so-far purely
theoretical) Majorana mode have created tremendous excitement \cite{R1} in the
general scientific community. Given the great significance of the possible
experimental discovery of the Majorana mode, it is therefore of utmost
importance to ensure that all aspects of the experimental discovery in Ref.[1]
are indeed consistent with the theoretical expectations and there are no loose
ends. This is particularly true in view of the facts that the experimental
observation precisely follows theoretical predictions for the existence of the
Majorana zero energy mode in semiconductor-superconductor hybrid structures
and that zero bias peaks could arise in superconductors and semiconductors
from a variety of physical effects such as Andreev or Shiba bound states,
Kondo resonances, etc. Our work presented in this paper aims at a qualitative
understanding of several interesting features of the experimental observation
to ensure that the observation is consistent with the expectations of the BdG
theory for the Majorana mode in semiconductor-superconductor structures.

In the current work, we theoretically investigate some of the peculiar aspects
of the experimental observations in Ref.[\onlinecite{1}] which were not
directly or explicitly predicted earlier \cite{2,3,4,5,6,7,8,9,10} in the
extensive theoretical work leading to the experimental observation of the
Majorana mode although some of the aspects we discuss in our work were
implicit in the theory. The experimental observation specifically concentrates
on the study of a zero-bias-conductance peak (ZBCP) in the current
(I)-voltage(V) differential conductance $\left(  \frac{dI}{dV}\right)  $ of
the tunneling spectroscopy of an InSb nanowire on superconducting NbTiN, which
manifests itself only in the presence of an external magnetic field
$B_{x}\left(  \gtrsim0.1\text{ T}\right)  $ oriented along the wire (taken to
be the $x$-axis in this paper). The existence of this ZBCP in
Ref.[\onlinecite{1}] for $B_{x}\neq0$ has been claimed to be the verification
of the theoretical prediction for the existence of the zero-energy Majorana
mode \cite{2,3,4,5} in the wire. The zero-energy Majorana mode exists as
localized quasiparticles at the ends of the superconducting wire and is a
manifestation of the system being in a chiral $p$-wave topological
superconducting (TS) phase as envisioned more than a decade ago \cite{9,10}.
The theory predicts \cite{2,3,4,5,6,7} the presence of the TS phase for
$V_{x}>V_{c}=\sqrt{\Delta^{2}+\mu^{2}}$, i.e. $B_{x}>B_{c}=V_{c}/g\mu_{B}$
with $V_{x}=g\mu_{B}B_{x}$ being the Zeeman field in the wire associated with
$B_{x},$ and $\Delta,\mu$ are the superconducting gap and the chemical
potential respectively in the wire. For $B_{x}<B_{c}$ (or $V_{x}<V_{c}$) the
system is in an ordinary non-topological (i.e. s-wave) superconducting phase
(NTS) which in the presence of the finite Zeeman splitting $V_{x}$ makes a
topological quantum phase transition \cite{2,3,4,5,6,7,8,9,10,11} to the TS
phase for $V_{x}>V_{c}$ (i.e. $B_{x}>B_{c}$). The TS phase has the Majorana
modes localized at the ends of the wire and the associated ZBCP at zero energy
in the middle of the superconducting gap. The NTS phase on the other hand has
no structure, except perhaps some Andreev bound states (ABS) at generic
non-zero energies, within the superconducting gap. The existence of a robust
ZBCP in the differential tunneling conductance has therefore been predicted
\cite{2,3,4,5,6,7,8,9,12} to be the necessary condition for the observation of
the Majorana mode, and its observation in Ref.[\onlinecite{1}] is an important
experimental milestone providing perhaps the first definitive signature for
the Majorana fermion in a solid state system.

Given the key importance of the subject matter, namely, the possible
experimental discovery of the emergent Majorana mode in the topological
superconductor system, it is somewhat disconcerting that some of the observed
experimental features are unexpected and somewhat inconsistent with the
existing theoretical predictions in the literature although most of the
findings in Ref.[\onlinecite{1}] are, in fact, completely consistent with the
theoretical predictions (e.g. the existence of ZBCP only above a critical
value of $B_{x}$). We concentrate in the current work on three features of the
experiment which, in our opinion, require special attention: $\left(
1\right)  $ the observed ZBCP is much (by more than an order of magnitude)
weaker than the predicted \cite{8,wimmer,9,12,ramon} canonical quantized value
of $2e^{2}/h$ expected for the Majorana zero energy mode; $\left(  2\right)  $
a peculiar and unexpected splitting of the ZBCP at high values of $V_{x}$ (for
$B_{x}\gtrsim0.5$ T) observed in Ref.[\onlinecite{1}]; $\left(  3\right)  $
the predicted behavior of the ZBCP in the presence of an additional transverse
component $V_{y}$ of the Zeeman field associated with an applied magnetic
field component $B_{y}\left(  =V_{y}/g\mu_{B}\right)  $ along the direction of
the spin-orbit coupling field $\left(  y\text{-axis}\right)  $ which is known
\cite{13} to be transverse to the length to the wire. Of the three issues
theoretically considered in this work, the first two are directly motivated by
the experimental data presented in Ref.[\onlinecite{1}] where a strongly
suppressed ZBCP (with a differential conductance value substantially below
$2e^{2/}h$ ) and a splitting of the ZBCP into a doublet at high values of
$B_{x}$ are both observed. Item $\left(  3\right)  $ in our work is alluded to
in Ref.[\onlinecite{1}], and our work here provides the numerical results for
the expected experimental observation when the applied in-plane $\vec{B}$
field is tilted at an angle $\theta$ to the wire length direction, i.e.
$\left(  B_{x},B_{y}\right)  =\left(  B\cos\theta,B\sin\theta\right)  ,$ which
gives $\left(  V_{x},V_{y}\right)  =\left(  g\mu_{B}B\cos\theta,g\mu_{B}%
B\sin\theta\right)  $ where $g,\mu_{B}$ are the Lande g-factor and the Bohr
magneton respectively.

The most important new qualitative feature of our theoretical work presented
in this work is considering realistic experimental systems, in particular,
finite wire lengths and finite temperatures as well as finite tunneling
barrier heights. We find that the finiteness of the nanowires is a fundamental
constraint in the ideal realization of the Majorana zero energy mode, and most
of our interesting and important results arise directly from our keeping wire
lengths finite as in the experimental systems. The reason for the qualitative
importance of the finite wire length is rather obvious. The external magnetic
field suppresses the superconducting gap, both in the NTS and the TS phase,
thus enhancing the coherence length which varies inversely as the gap energy.
When the enhanced coherence length becomes comparable to the wire length, the
two end Majorana modes start 'seeing' each other, leading to an energy
splitting. This effect turns out to be of qualitative importance, affecting
for example the conductance quantization of the Majorana ZBCP even at $T=0$ in
contrast to the ideal infinite wire case where the ZBCP is always quantized at
$2e^{2}/h$ in the $T=0$ limit.

The quantization of the ZBCP predicted for Majorana fermions
\cite{8,wimmer,9,12} is a result that is valid only in the zero-temperature
limit. In contrast, the high-temperature limit has a more conventional
resonant scattering Fano-form ~\cite{bolech_demler,wimmer,4} with a height
proportional to $\Gamma/k_{B}T$ and a width proportional to the thermal energy
$k_{B}T$, for $k_{B}T\gg\Gamma$ with $\Gamma/\hbar$ being the tunneling rate
between the Majorana bound state and the lead. The tunneling rate
$\Gamma/\hbar$ depends on the transparency of the tunneling barrier, and is
not easily experimentally controllable-- in fact, the barrier transparency and
hence the tunneling rate is simply unknown in the experimental situation. In
the part of this work which focuses on Item $\left(  1\right)  $, we show how
the tunneling conductance crosses over from the low-temperature to the
high-temperature limit and establish that for reasonable parameters, it is
indeed possible to have a dramatic suppression of the ZBCP as seen in
Ref.[\onlinecite{1}]. In addition, from our numerical calculations for the
realistic parameters, we have found that even at $T=0$, the ZBCP can be
suppressed below its quantized value for sufficiently small tunneling rate
$\Gamma$ because of finite size effects. In particular, such a $T=0$
suppression of ZBCP happens when $\Gamma$ becomes comparable to the splitting
between the end Majorana fermions, which may be the case for the few micron
long wires in Ref.[\onlinecite{1}] together with the tunneling rates $\Gamma$
inferred from the measured conductance. The splitting between end Majoranas is
invariably present in real wires where the two end Majoranas have some finite
overlap, leading to a lifting of their precise zero energy status. The
splitting of the ZBCP as a function of $V_{x}$, which is discussed as part of
Item $\left(  2\right)  $ in this paper is a finite size effect, which is
likely to be relevant for the experiments in Ref.[\onlinecite{1}], but as far
as we are aware has not been discussed in the literature except in the
idealized situation\cite{meng}. The predictions in the literature, which are
restricted to infinite wires, show that the Majorana fermion must be robust
for large Zeeman fields in the case of narrow wires where the inter-sub band
spacing is much larger than the Zeeman splitting. The numerical results
presented in this paper show that for finite wires, even in the narrow wire
limit, the ZBCP is split for large $V_{x}$. The splitting of the ZBCP arises
from overlap of the Majorana fermion wave-functions as has been previously
discussed in the context of $p$-wave superconductors\cite{meng}. To the best
of our knowledge, our work is the only existing result in the literature for
the Majorana splitting in finite wires in the presence of spin-orbit coupling
and Zeeman splitting. Finally in the context of Item $\left(  3\right)  $, we
discuss the effect of the angle of the Zeeman potential on the ZBCP.
Consistent with previous theoretical work \cite{4}, which shows that the
proximity-induced quasiparticle gap vanishes in the wire for $V_{y}>\Delta$,
we find that the ZBCP vanishes above a threshold value of $V_{y}$. 

We emphasize that our goal in this work is not to develop a new theory for the
Majorana modes in the semiconductor nanowires, but to extend and generalize
the existing theory\cite{2,3,4} leading to the prediction of the Majorana mode
in semiconductor hybrid structures, to finite wire lengths, temperatures, and
barrier heights as well as to large Zeeman splitting (along arbitrary
directions also) to see whether the features arising out of the standard
theory are consistent with the observations of Ref.[\onlinecite{1}]. 
Of course, theory by itself cannot answer the question whether the observations in
 Ref.[\onlinecite{1}] are really the isolated Majorana modes predicted by theory 
\cite{2,3,4} or are some disorder or multi-mode wire effect. Instead our goal 
is to show that some of the qualitative features of the experiment, which are not 
immediately consistent with the results explicitly predicted in previous works \cite{2,3,4}, 
are indeed consistent with the experiment. 
%\sout{The standard theory so far has understandably concentrated on the ideal situation
%with infinite wire length, near perfect or infinitesimal transmission, zero
%temperature, though}  
therefore, it is important to discern which features of the
experimental data are in direct agreement with the theory extended to include
finite wire length, finite temperature, finite barrier height, and large
Zeeman field. Such a comparison, as carried out in this work, between the
extended standard theory and the data of Ref.[\onlinecite{1}] will help
indicate whether the experiment\cite{1} really is consistent with a Majorana
interpretation and also whether new ideas are essential to understand some
features of the data. 
As discussed in more detail in Sec. VI, we intentionally avoid discussing issues such as 
disorder and the multi-band effects, since these effects would complicate the conclusions 
by introducing more unknowns into the theory. Moreover, as elaborated further in Sec. VI, 
the experimental data in Ref.[\onlinecite{1}] 
appears to use wires with relatively long mean-free paths (300 nm) and have large 
sub-band spacings so that such effects are unlikely to change our conclusions in a qualitative way.
In this context, it is also important to emphasize that
the finite wire length automatically implies a crossover in the behavior of
the Majorana mode when the magnetic field induced suppression of the
superconducting induced gap leads to the coherence length becoming comparable
to or larger than the wire length. This physics is qualitatively new in our
work (although it is implicit in the earlier works) because a finite wire
length allows the two Majorana modes to overlap with each other leading to a
splitting which is by definition impossible in an infinite wire at any
magnetic field. Since the experiments are always done in finite wires, our
work provides a crucial extension of the standard theory in order to
understand or interpret the experimental results even at a qualitative level.

\section{\bigskip Theory}

The physical system \cite{2,3,4} for studying Majorana fermions includes a
strongly spin-orbit coupled semiconductor (SM), proximity-coupled to an
$s$-wave superconductor (SC) and imposed to a Zeeman field. Without loss of
generality, we consider a finite 1D SM nanowire along the $\hat{x}$ direction,
the spin-orbit interaction $\alpha_{R}$, being along the $y$ axis, and a
Zeeman field $\vec{V}=\left(  V_{x},V_{y}\right)  $. Also the wire is in
contact with a superconductor, with proximity induced pairing strength
$\Delta.$ The continuous BdG Hamiltonian for the system is
\begin{equation}
H=\left(  -\frac{\hbar^{2}}{2m}\partial_{x}^{2}-\mu\right)  \tau_{z}%
+V_{x}\sigma_{x}+V_{y}\sigma_{y}+i\alpha_{R}\partial_{x}\sigma_{y}\tau
_{z}+\Delta\tau_{x}.\label{1}%
\end{equation}
$\mu$ is the chemical potential. The Pauli matrices $\sigma,\tau$ operate in
spin and particle-hole space, respectively. For numerical convenience, it is
standard to study the BdG equation in a discrete lattice tight-binding
approximation with no loss of generality and with fewer unknown parameters.
Under the lattice approximation, we can map $\left(  \ref{1}\right)  $ to a
tight-binding model%
\begin{align}
H &  =\sum_{i,j,\sigma}t_{ij}c_{i\sigma}^{\dagger}c_{j\sigma}-\sum_{i,\sigma
}\mu_{i}c_{i\sigma}^{\dagger}c_{i\sigma}\label{2}\\
&  +\sum_{i,\sigma\sigma^{\prime}}\frac{\alpha_{i}}{2}\left[  c_{i+1\sigma
}^{\dagger}\left(  i\sigma_{y}\right)  _{\sigma\sigma^{\prime}}c_{i\sigma
^{\prime}}+H.c.\right]  \nonumber\\
&  +\sum_{i,\sigma}c_{i\sigma}^{\dagger}\left(  V_{x}\left(  i\right)
\sigma_{x}+V_{y}\left(  i\right)  \sigma_{y}\right)  _{\sigma\sigma^{\prime}%
}c_{i\sigma^{\prime}}\nonumber\\
&  +\sum_{i}\Delta_{i}\left(  c_{i\uparrow}^{\dagger}c_{i\downarrow}^{\dagger
}+H.c.\right)  \tau_{x}\nonumber
\end{align}
The first contribution describes hopping or the kinetic energy along the
length (i.e. the x-axis) of the wire, the second term represents the Rashba
spin-orbit interaction, the third term is the Zeeman field and the last term
shows the proximity-induced pairing term. $c_{i\sigma}^{\dagger},c_{i\sigma}$
denote electron creation and annihilation operators, respectively.\ We include
only nearest-neighbor hopping with $t_{ij}=-t_{0}$ and also include an on-site
contribute $t_{ii}=2t_{0}$ that shifts the bottom of the energy spectrum to
zero energy. The chemical potential $\mu$ is calculated from the bottom of the
band. In the long wavelength limit, the tight binding model reduces to the
continuum Hamiltonian $\left(  \ref{1}\right)  $ with $t_{0}=\hbar
^{2}/2m^{\ast}a^{2}$ and Rashba spin-orbit coupling $\alpha_{R}=\alpha a$ with
lattice constant $a.$ In the numerical calculations we use a set of parameters
consistent with the properties of InSb, as in Ref.[1], and choose the
effective mass $m^{\ast}=0.015m_{e},$spin-orbit coupling $\alpha_{R}=0.2$
$meV\mathring{A},$ and $g\mu_{B,InSb}=1.5$ $meV/T$. We also use $\Delta=0.5$
$meV$ and the length of the wire to be $4.5$ $\mu m$. These parameters are
roughly consistent with the experimental conditions of Ref.[\onlinecite{1}]
although we are not interested in fine-tuning parameters for quantitative
agreement with the data because there are far too many unknown parameters in
the experiment. We choose the tight binding numerical lattice parameter
$a=15\text{ nm},$ which is chosen so that the band-width $2t_{0}\gg
V,\Delta,\mu$ and thus the tight binding approximation itself would not
introduce any artifacts into our results. The length of the wire $L=4.5\,\mu
m$ then corresponds to $N=300$ sites. We mention here that the SC proximity
effect has now been observed by several groups in the SC/SM hybrid systems
including both InSb nanowires \cite{1,nilsson,rokhinsonprivate} and InAs
nanowires\cite{doh,marcusprivate}. Thus, the effective model given by
Eq.\ref{2}, which is the starting point for our theory, is an appropriate
model for the SM nanowire in the presence of spin-orbit coupling, Zeeman
splitting, and proximity-induced superconductivity.

To calculate the differential conductance measured from tunneling into the end
of the superconducting nanowire, we have to study the current flowing into the
wire contacted through a barrier region at one end with a lead by using the
Blonders-Tinkham-Klawijk (BTK) formalism.\cite{16} The main idea is to get the
reflection and transmission coefficients by solving a BdG eigen-equation with
some initial conditions. The current and conductance can be expressed in terms
of these coefficients. Given that the experiment involves many unknown
parameters controlling many interfaces (e.g. SC/SM, tunneling barrier/SM,
various gates to control the barriers and chemical potential), our goal is to
utilize the simplest possible model with the fewest number of parameters which
would be capable of capturing the underlying qualitative physics of the
experiment in finite wires. The BdG-BTK formalism provides the simplest
qualitative basis for the theoretical modelling of the experimental system in Ref.[1].

More precisely, we start with the tight-binding Hamiltonian $\left(
\ref{2}\right)  $ with open boundary conditions so that the part
$i=0,1,...m_{L}$ are in the lead with no superconductivity and the sites
$i=m_{L}+1,...,N$ are in the superconducting nanowire. The barrier region is
modeled as a variation in the local chemical potential $\mu_{i}\rightarrow
\mu_{i}-U$ for the sites $i=m_{L}-2,\dots,m_{L}+2$, where $U$ is the tunnel
barrier height. To calculate the reflection and transmission coefficients at
energy $E,$ we note that since the lead is normal (without spin-orbit
coupling), the incoming mode can be taken to be purely electron-like. The
reflected amplitudes can be normal $r_{N,\sigma,\sigma^{\prime}}$ or anomalous
$r_{A,\sigma,\sigma^{\prime}}.$ Here $\sigma$ is the spin of incoming electron
and $\sigma^{\prime}$ is the spin of the reflected electron and hole. The BdG
equations that determine the reflected coefficients are%
\begin{align*}
&  \sum_{n=0,N}\left(  H_{m,n}-E\delta_{m,n}\right)  \Psi_{n}=0\text{ for
}m=1,..N\\
&  \Psi_{n=0}=\left(
\begin{array}
[c]{c}%
\delta_{\sigma,\uparrow}\\
\delta_{\sigma,\downarrow}\\
0\\
0
\end{array}
\right)  +\left(
\begin{array}
[c]{c}%
r_{N,\sigma,\uparrow}\\
r_{N,\sigma,\downarrow}\\
r_{A,\sigma,\uparrow}\\
r_{A,\sigma,\downarrow}%
\end{array}
\right)  \\
&  \Psi_{n=1}=\left(
\begin{array}
[c]{c}%
\delta_{\sigma,\uparrow}\\
\delta_{\sigma,\downarrow}\\
0\\
0
\end{array}
\right)  e^{ik_{e}a}+\left(
\begin{array}
[c]{c}%
r_{N,\sigma,\uparrow}\\
r_{N,\sigma,\downarrow}\\
0\\
0
\end{array}
\right)  e^{-ik_{e}a}\\
&  +\left(
\begin{array}
[c]{c}%
0\\
0\\
r_{A,\sigma,\uparrow}\\
r_{A,\sigma,\downarrow}%
\end{array}
\right)  e^{ik_{h}a}.
\end{align*}
The first equation is the BdG equation in the superconducting wire, while the latter two equations 
express the wave-functions in the lead in terms of the reflection coefficients.
These equations need to be solved for both $\sigma=\uparrow,\downarrow.$ In
the above $k_{e}$ is the wave-vector in the lead at energy $E$ so that
\[
\frac{k_{e}^{2}}{2m}-\mu_{lead,\sigma}=E
\]
where $\mu_{Lead}$ is the chemical potential of the lead, while $k_{h}$
satisfies%
\[
\frac{k_{h}^{2}}{2m}-\mu_{lead,\sigma}=-E.
\]
The voltage-bias $V$ of the lead determines the occupancy of the incident
electrons. Electrons are incident on the superconductor from energy
$E=-\mu_{lead}$ to $E=V.$ The states with normal reflection do not contribute
to the current. Thus, the current will be%
\[
I=\sum_{\sigma}\int_{-\mu_{lead,\sigma}}^{V}dE\sum_{\sigma^{\prime}}\left\vert
r_{A,\sigma,\sigma^{\prime}}\left(  E\right)  \right\vert ^{2}%
\]
which implies a conductance (in unit of $2e^{2}/h)$%
\[
\frac{dI}{dV}=\sum_{\sigma,\sigma^{\prime}}\left\vert r_{A\sigma\sigma
^{\prime}}\left(  E=V\right)  \right\vert ^{2}.
\]

Technically for negative bias voltages $V<0,$ one needs to consider holes
below the fermi-energy incident from the right with energy $E=-V>0$ and then
Andreev reflected at the interface and becoming electrons. However, such
processes are related by particle-hole symmetry to Andreev reflection of
negative energy electrons. Therefore, one does not need to calculate the
Andreev reflection process separately since it is automatically incorporated
in our BdG-BTK formalism.

To generalize to finite temperatures all one needs to do is to broaden the
conductance with the derivative of the Fermi-function in the usual manner%
\[
G\left(  V,T\right)  =\int d\varepsilon G_{0}\left(  \varepsilon\right)
\frac{1}{4T\cosh^{2}\left(  \left(  V-\varepsilon\right)  /2T\right)  }%
\]
with $G_{0}\left(  \varepsilon\right)  $ the zero temperature conductance at
energy $\varepsilon.$ \begin{comment}
We also include in our theory the effect of the finite
tunnel barrier using the barrier potential parameter $U$ which \ goes into the
theory in a straightforward manner. Specifically we model the tunneling
barrier in the system by an onsite potential $U$ at the interface between the
lead and the superconducting wire.
\end{comment}

We present our numerical results for the calculated tunneling conductance as a
function of bias voltage and other relevant experimental tuning parameters in
Figs.1-5 to be compared with the experimental data of Ref.[1]. We first
mention the fact that the (Majorana) properties of the real experimental
system depend on (at least) ten independent parameters (many of which are
unknown). These parameters include temperature, tunneling barrier (which in
all likelihood depends on several unknown parameters determined by the details
of the interface and the control gates), Zeeman fields ($V_{x}$ and $V_{y}$),
spin-orbit coupling, chemical potential, the induced superconducting gap
(which in turn depends on several parameters such as the
semiconductor-superconductor hopping amplitude, disorder, and the parent gap
in NbTiN), the parameters defining the 1D confinement in the wire (which
requires at least four independent parameters for confinement along $y$ and
$z$ directions), wire length $\left(  L\right)  $ along the wire, and disorder
(which by itself would necessitate several independent parameters for its
description since in principle there could be long-ranged and short-ranged
random impurities in the wire as well as interfacial roughness at the
semiconductor-superconductor interface). No meaningful theory, beyond mere
data fitting, can of course attempt to include all these parameters. In the
current work we are interested in the fundamental question of whether a
minimal theoretical model can capture the basic qualitative findings of
Ref.[\onlinecite{1}], and as such we ignore all the complications,
concentrating on the single subband 1D model in the absence of disorder within
a tight-binding BdG-BTK formalism. Our work should be construed as a zeroth
order effective model of the experiment which should be the starting point for
future quantitative and realistic theoretical studies when the details of the
experimental parameters become available.
\begin{comment}Multisubband occupancy\cite{6,7} and disorder \cite{7,12} effects,
which have been considered elsewhere in the literature, do not change any of
our conclusions qualitatively as long as we are in the narrow wire limit
where inter-sub-band spacing is larger than a few meV. We have checked this statement numerically by
including disorder and multisubband occupancy in our calculations.%
\end{comment}

%

%TCIMACRO{\FRAME{ftbpFU}{3.1669in}{2.2416in}{0pt}{\Qcb{(Two Left Panels) The
%differential conductance for a fixed Zeeman potential $\QTR{bf}{V}=\left(
%1,0\right)  $ meV and different tunneling barrier $U=38$ and $42$ meV. The
%green dotted, red dashed and orange lines denote different temperature
%$0,60,120$ mK, respectively. (Two Right Panels) Zero bias conductance peak as
%a function of temperature for Zeeman potential $\QTR{bf}{V}=\left(
%1,0\right)  $ meV and different tunneling barrier $U=38$ and $42$ meV,
%respectively. The peak decreases monotonously with temperature. ($L=4.5\mu m,$
%spin-orbit coupling $\alpha=0.2$ meV\AA , induced pairing potential
%$\Delta=0.5$ meV)}}{\Qlb{fig1}}{fig1.eps}%
%{\special{ language "Scientific Word";  type "GRAPHIC";  display "USEDEF";
%valid_file "F";  width 3.1669in;  height 2.2416in;  depth 0pt;
%original-width 14.2936in;  original-height 8.8194in;  cropleft "0";
%filename 'fig1.eps';file-properties "XNPEU";}} }%
%BeginExpansion
\begin{figure}
[ptb]
\begin{center}
\includegraphics[
height=2.2416in,
width=3.1669in
]%
{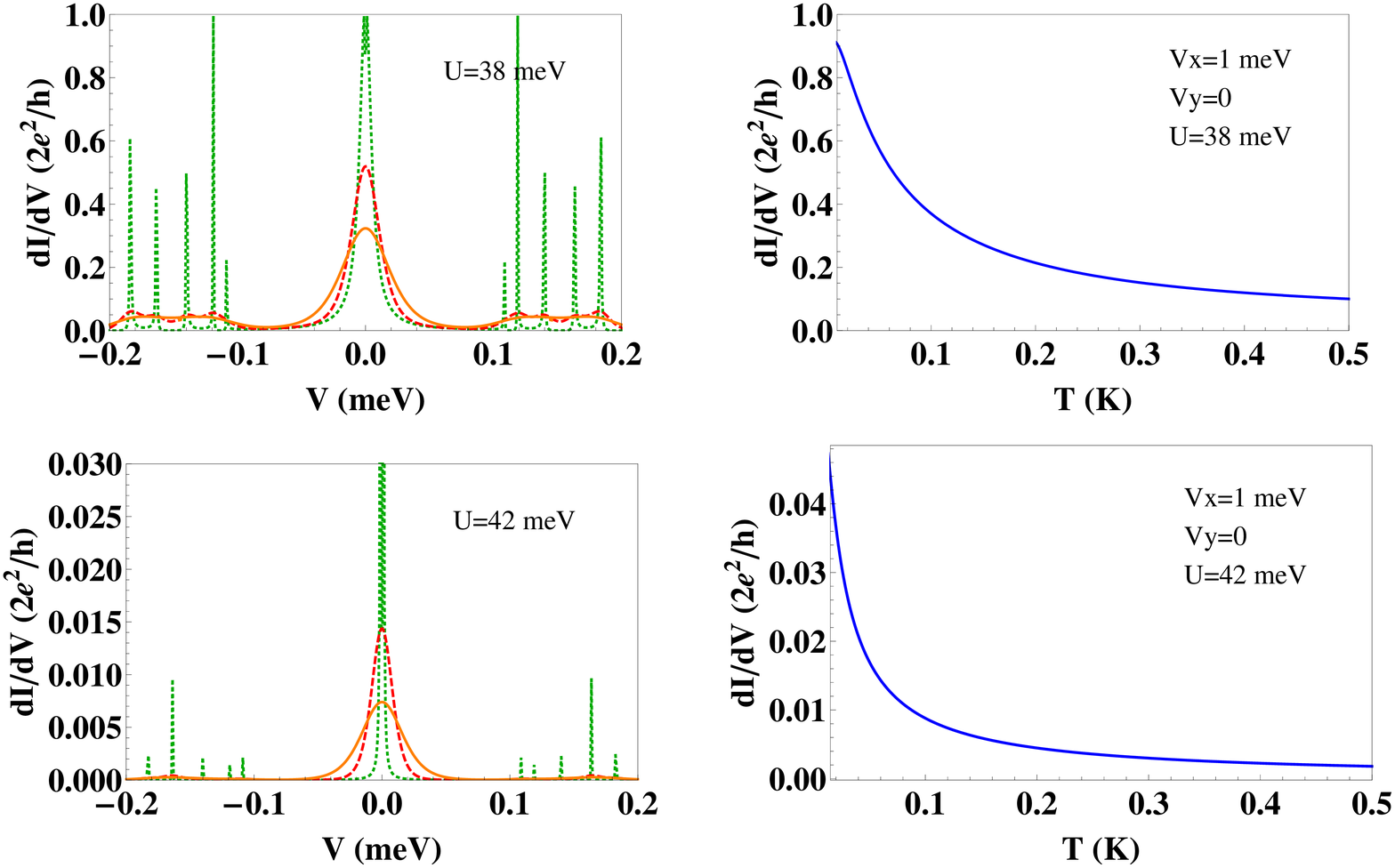}%
\caption{(Two Left Panels) The differential conductance for a fixed Zeeman
potential $\mathbf{V}=\left(  1,0\right)  $ meV and different tunneling
barrier $U=38$ and $42$ meV. The green dotted, red dashed and orange lines
denote different temperature $0,60,120$ mK, respectively. (Two Right Panels)
Zero bias conductance peak as a function of temperature for Zeeman potential
$\mathbf{V}=\left(  1,0\right)  $ meV and different tunneling barrier $U=38$
and $42$ meV, respectively. The peak decreases monotonously with temperature.
($L=4.5\mu m,$ spin-orbit coupling $\alpha=0.2$ meV\AA , induced pairing
potential $\Delta=0.5$ meV)}%
\label{fig1}%
\end{center}
\end{figure}
%EndExpansion

\section{\bigskip Results}

In Fig.\ref{fig1} we show our calculated differential conductance $dI/dV$ as a
function of the tunneling bias voltage $V$ (which should not be confused with
the Zeeman fields $V_{x}$,$V_{y}$) for two different tunnel barriers and three
different temperatures for $V_{x}=1$ meV and $V_{y}=0.$ (This choice of
$\mathbf{V=}\left(  V_{x},V_{y}\right)  $ guarantees that the system is in the
TS phase satisfying $V_{x}>\sqrt{\Delta^{2}+\mu^{2}}$ for our choice of system
parameters corresponding to Ref.[1]$.$) In the third panel of Fig.\ref{fig1},
we depict $\frac{dI}{dV}$ of the ZBCP itself as a function of $T$ for two
values of $U.$ These results manifestly establish that the canonical quantized
value of $2e^{2}/h$ is clearly an unphysical theoretical limit achievable only
as $T\rightarrow0$ (and for low values of $U$). For reasonable values of $U$
and $T$, our calculated value of ZBCP in Fig.\ref{fig1} could easily be one to
two orders of magnitude smaller than $2e^{2}/h,$ thus providing a satisfactory
probable explanation for the weak strength of the ZBCP observed in
Ref.[\onlinecite{1}]. We note that in our model, low values of U correspond to
the more transparent barrier with higher conductance values-- our numerical
results clearly indicate that the ideal Majorana conductance of $2e^{2}/h$ is
unphysical even in the extreme low-temperature limit since the tunnel barrier
U is never likely to be very small (or equivalently the transparency to the Majorana 
is not likely to be large).
 Our work leads immediately to the somewhat disappointing conclusion that the
 full Majorana spectral strength may never be
achievable in the standard experimental set ups for measuring dI/dV since it
would be very hard to get to the ideal limit of both $T$, $U$ being very small
(compared with the induced topological gap).  
Without our numerical results,
this would not be apparent.
Since the conductance calculated for the larger barrier height in Fig.~\ref{fig1} 
is found to be significantly below the quantized conductance even at $T=0$, 
it is clear that the suppression of conductance in Fig.~\ref{fig1} is not purely 
the previously studied finite temperature effect \cite{8,bolech_demler}.
 Of course, as mentioned in the Introduction, this
finding of ours, namely that the Majorana tunneling differential conductance
is not quantized even at $T=0$ for finite $U$ is valid only for finite wires
since it is well-established that in the thermodynamic limit at $T=0$ the
Majorana conductance is necessarily quantized to a value of $2e^{2}/h$. The
key here is that in finite wires the two end Majoranas always overlap, and one
must have the tunneling barrier low enough for this overlap to be unimportant
in determining the conductance quantization-- thus for very long wires the
conductance quantization will be valid up to very large values of $U$, but for
short wires, the value of $U$ must be quite low. Our finding of the strong
suppression of the ZBCP in short wires even in the $T=0$ limit due to the
tunnel barrier effect (coupled with the Majorana overlap) is a new and
somewhat unexpected result, which is consistent with the experimental finding
of a saturation of the ZBCP at low temperatures\cite{1}.
While splitting of the MFs  resulting from the finite length 
of the wire is expected to  suppress the ZBCP because splitting of the MFs removes 
the zero-energy state responsible for the ZBCP, one would expect such splitting 
to result in a split ZBCP rather than one that is simply reduced in magnitude.
A split ZBCP should be discernible in experiments at low temperatures and 
in this case the splitting of the MFs would be easy to determine from the 
experimental results. Similarly, a reduction of the ZBCP from finite temperature 
would lead to a temperature dependenct ZBCP, whose width is comparable to $k_B T$.
 Our results plotted in  Fig.~\ref{fig1} 
show that for an intermediate regime of the barrier height, $U$, the finite 
overlap of the MFs can suppress the ZBCP without leading to an observable 
splitting of the ZBCP. 
 Extremely large barrier heights $U$ lead to peak width, 
which is dominated by temperature, and a temperature-dependent ZBCP height. 
Very low temperatures $T$ and high barrier heights could in principle lead 
to a regime where a split conductance is observed. 
On the other hand, low temperatures and a barrier with a transparency smaller than 
the finite size splitting would result in a ZBCP, which is at the same time 
independent of temperature and also have a ZBCP height that is significantly 
smaller than $ 2 e^2/h$. Our results show that finite size effects 
can suppress a ZBCP significantly below $2 e^2/h$, without the peak 
being temperature-independent or split.

%

%TCIMACRO{\FRAME{ftbpFU}{3.2828in}{3.1618in}{0pt}{\Qcb{The differential
%conductance for a fixed $V_{y}=0$ meV and different $V_{x}=0,0.5,1,2,3,4$ meV.
%The red solid and blue dashed lines denote temperatures $T=0$ and $60$ mK,
%respectively. At $V_{x}=1$ meV, the system is in the topological phase with
%quantized ZBCP. With larger $V_{x},$ it will reduce the gap, lead to stronger
%overlap between MFs and result in suppression of ZBCP. Also the finite
%temperature will decrease ZBCP. ($U=38$ meV, $L=4.5\mu m,$ spin-orbit coupling
%$\alpha=0.2$ meV\AA , induced pairing potential $\Delta=0.5$ meV)}}%
%{\Qlb{fig2}}{fig2.eps}{\special{ language "Scientific Word";  type "GRAPHIC";
%display "USEDEF";  valid_file "F";  width 3.2828in;  height 3.1618in;
%depth 0pt;  original-width 13.3233in;  original-height 12.5562in;
%cropleft "0";  croptop "1";  cropright "1";  cropbottom "0";
%filename 'fig2.eps';file-properties "XNPEU";}} }%
%BeginExpansion
\begin{figure}
[ptb]
\begin{center}
\includegraphics[
height=3.1618in,
width=3.2828in
]%
{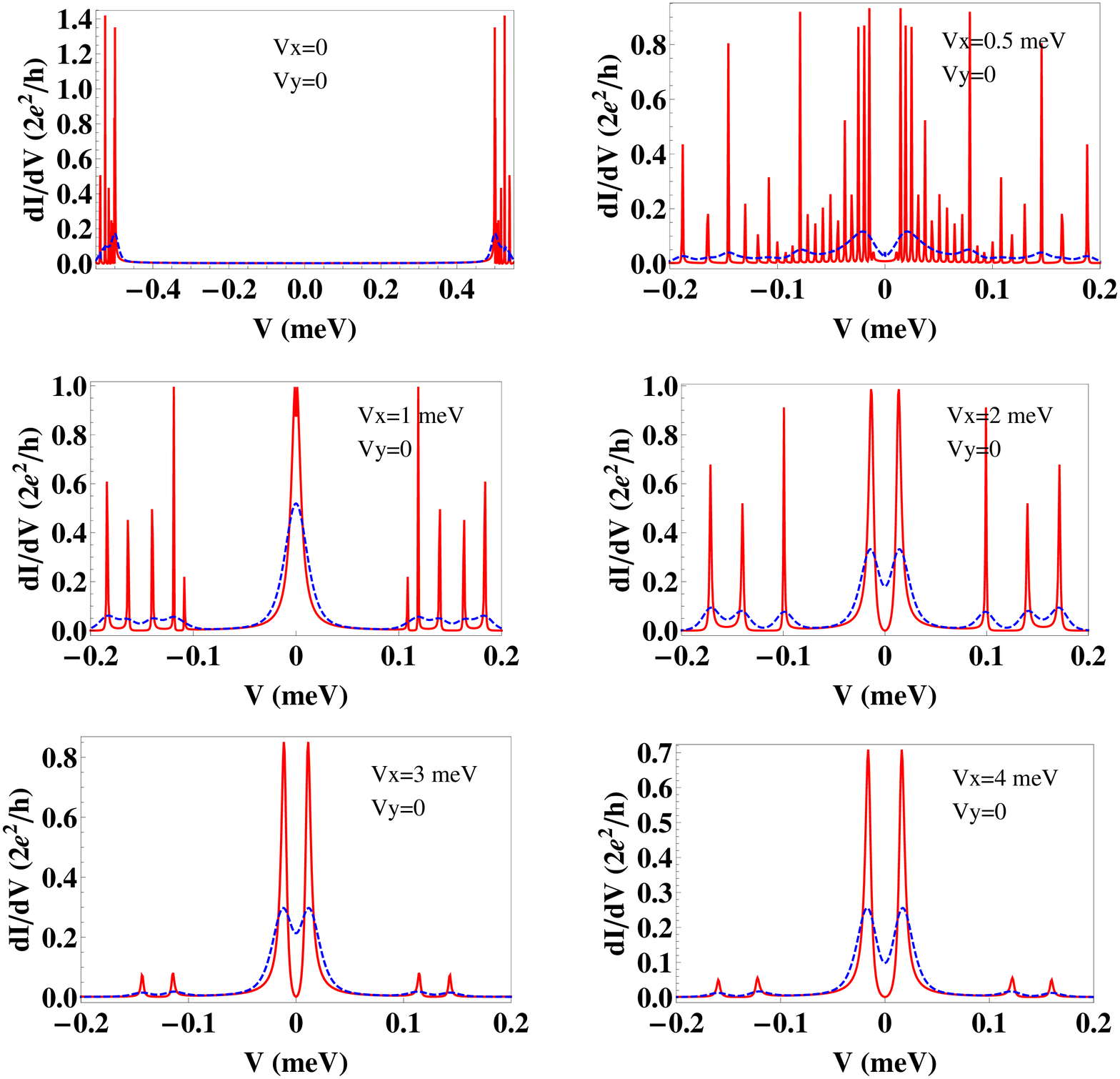}%
\caption{The differential conductance for a fixed $V_{y}=0$ meV and different
$V_{x}=0,0.5,1,2,3,4$ meV. The red solid and blue dashed lines denote
temperatures $T=0$ and $60$ mK, respectively. At $V_{x}=1$ meV, the system is
in the topological phase with quantized ZBCP.
For length $L=4.5\mu m,$ the Zeeman splitting $V_{x}\gg 1$ meV reduces the
 gap, and leads to a stronger overlap between the MFs that results in a splitting of the ZBCP peaks.
Additionally the finite temperature will decrease ZBCP. (Parameters used in the plot: $U=38$ meV, $L=4.5\mu m,$
spin-orbit coupling $\alpha=0.2$ meV\AA , induced pairing potential
$\Delta=0.5$ meV)}
\label{fig2}%
\end{center}
\end{figure}
%EndExpansion

In Fig.\ref{fig2}, we show our calculated magnetic field or Zeeman splitting
$V_{x}$ dependence ($V_{y}=0$) of $dI/dV$ for two fixed temperatures keeping
all other parameters fixed. The interesting result here in partial agreement
with Ref.[\onlinecite{1}] is the splitting of the ZBCP for large ($V_{x}\geq2$
meV) values of the Zeeman splitting. This ZBCP splitting arises from the wire
length ($L=4.5$ $\mu m$) being finite (both in our simulations and in
Ref.[\onlinecite{1}]), which leads to the possibility of the two Majorana
modes localized at the two ends of the wire to hybridize \cite{meng,14}
causing the splitting of the ZBCP. (The double-peak structure of the ZBCP is
apparent in Ref.[1] for the applied magnetic field above $\sim0.48$ T where
two peaks around zero bias can be seen in the data.) The Majorana
hybridization effect (and consequently the splitting of the ZBCP) is
exponentially suppressed for smaller values of $V_{x}\left(  >V_{c}\right)  $
still within the TS phase since the superconducting gap is large. With
increasing $V_{x},$ the gap is eventually suppressed as $V_{x}^{-1}$, which
increases the coherence length, leading to an effective overlap between the
Majorana modes localized at the two ends of the wire. There is nothing
mysterious about the splitting of the ZBCP at high values of $V_{x}$ where the
induced gap is small in the TS phase; this is expected-- the important point
is our finding that this happens in the same range of $V_{x}$ values in our
theoretical modeling as it does in the experiment. Interestingly for the
parameters of the problem in Fig.\ref{fig2}, we find the Zeeman induced
splitting of the ZBCP to be only weakly dependent on Zeeman field. In
particular, consistent with the data in Ref.[\onlinecite{1}] our calculated
ZBCP splitting $(\sim0.02$ meV$)$ in Fig.2 is much smaller than the applied
Zeeman field $(V_{x}\sim2-4$ meV$)$ causing this splitting-- this implies that
the ZBCP splitting is not a trivial spin splitting either in our theory or in
the experiment. As already mentioned above, the ZBCP splitting in the theory
has its origin in the splitting of the Majorana zero energy mode due to the
finite overlap of the two end Majorana localized wavefunctions overlapping due
to the finite length and the high field situation, as predicted originally in
Ref.[\onlinecite{meng}]. We emphasize that the ZBCP splitting depends on
increasing $V_{x}$ in a nontrivial manner and is not simply a Zeeman splitting
going as linear in B, which serves to distinguish the high-field Majorana
splitting from any run-of-the-mill Zeeman splitting arising in the
experimental situation.

The results presented in Fig.2 clearly indicate that the ZBCP splitting (i.e.
the Majorana splitting in finite wires) as a function of the Zeeman splitting
is a non-monotonic function of the Zeeman field $V_{x}$, as was first
predicted theoretically in Refs.[\onlinecite{meng}]and [\onlinecite{14}] in a
slightly different context. In fact, the splitting oscillates with the
magnetic field $V_{x}$ since it is an oscillatory function of the
superconducting coherence length\cite{meng,14}. This feature was also observed
in the numerical work of Ref.[\onlinecite{ramon}]. We discuss this Majorana
splitting in more details in discussing Fig.6 below, but the oscillatory
behavior is already apparent in the results for the various $V_{x}$ values
shown in Fig.2 for smaller values of $V_{x}$. 
%\sout{It is, however, interesting to
%note that the ZBCP itself does not oscillate with $V_{x}$ at large values of
%$V_{x}$, essentially saturating to a $V_{x}$ independent value eventually. }

The observed high-field $\left(  \geq0.48\text{ T}\right)  $ splitting of the
ZBCP in Ref.[\onlinecite{1}] could thus probably arise from a finite wire
length effect in the high field regime. Of course a quantitative comparison
with experiment of the exact nature of the splitting of the ZBCP would require
a systematic determination of the disorder and field configuration of the
experiment. Also, the experimental data in Ref.[1] unfortunately end around
$0.5$ T value of the magnetic field, and it is therefore not possible right
now to quantitatively compare the field dependence of our numerical finding of
the ZBCP splitting arising from the Majorana overlap with the experimental
observation. Hopefully, our theory will motivate further high-field studies so
that the dependence of the ZBCP splitting on $V_{x}$ can be compared between
experiment and theory.

It is known\cite{meng,14} that the Majorana splitting due to the overlap of
two Majorana modes decays exponentially with their separation with the
characteristic decay length being the superconducting coherence length. The
main physics of the enhanced splitting with increasing $V_{x}$ found in Fig.2
is thus the physics of the enhancement of the coherence length due to the gap
suppression by the external magnetic field. Such a splitting with increasing
$V_{x}$ is thus expected at high field values, but the important thing to note
is that the splitting is consistent with the experimentally observed ZBCP
splitting in the data of Ref.[1], lending credence to the claim that Ref.[1]
is really exploring Majorana physics. Note that in an infinite system, such a
splitting would never arise and the ZBCP will never split no matter how large
$V_{x}$ is. It is also known that in addition to the decay of the MF splitting
in finite systems, characterized by the separation between the MFs normalized
to the superconducting coherence length, the splitting also shows a
characteristic oscillatory behavior as a function of the coherence length
itself. In Fig.2, $V_{x}$ is kept fixed in each panel, so these oscillations
are not apparent, but we discuss these MF splitting oscillations as a function
of $V_{x}$ (through the enhancement of the coherence length by $V_{x}$) in the
Sec. V of the paper (see, e.g., Fig.6(d) which explicitly shows the
oscillation as a function of $V_{x}$).%

%TCIMACRO{\FRAME{ftbpFU}{3.3434in}{2.2295in}{0pt}{\Qcb{The differential
%conductance for a fixed $V_{x}=1$ meV and different $V_{y}=0.1,0.4,0.6$ and
%$0.8$ meV. The red solid and blue dashed denote temperatures $T=0$ and $60$
%mK, respectively. At $V_{x}=1$ meV, the system is in the topological phase
%with quantized ZBCP. $V_{y}$ will reduce the quasi-particle gap and hence
%suppress the ZBCP. Finite temperature also reduces the ZBCP. ($U=38$ meV,
%$L=4.5\mu m,$ spin-orbit coupling $\alpha=0.2$ meV\AA , induced pairing
%potential $\Delta=0.5$ meV)}}{\Qlb{fig3}}{fig3.eps}%
%width 3.3434in;  height 2.2295in;  depth 0pt;  original-width 12.5588in;
%original-height 7.7513in;  cropleft "0";  croptop "1";  cropright "1";
%cropbottom "0";  filename 'fig3.eps';file-properties "XNPEU";}} }%
%BeginExpansion
\begin{figure}
[ptb]
\begin{center}
\includegraphics[
height=2.2295in,
width=3.3434in
]%
{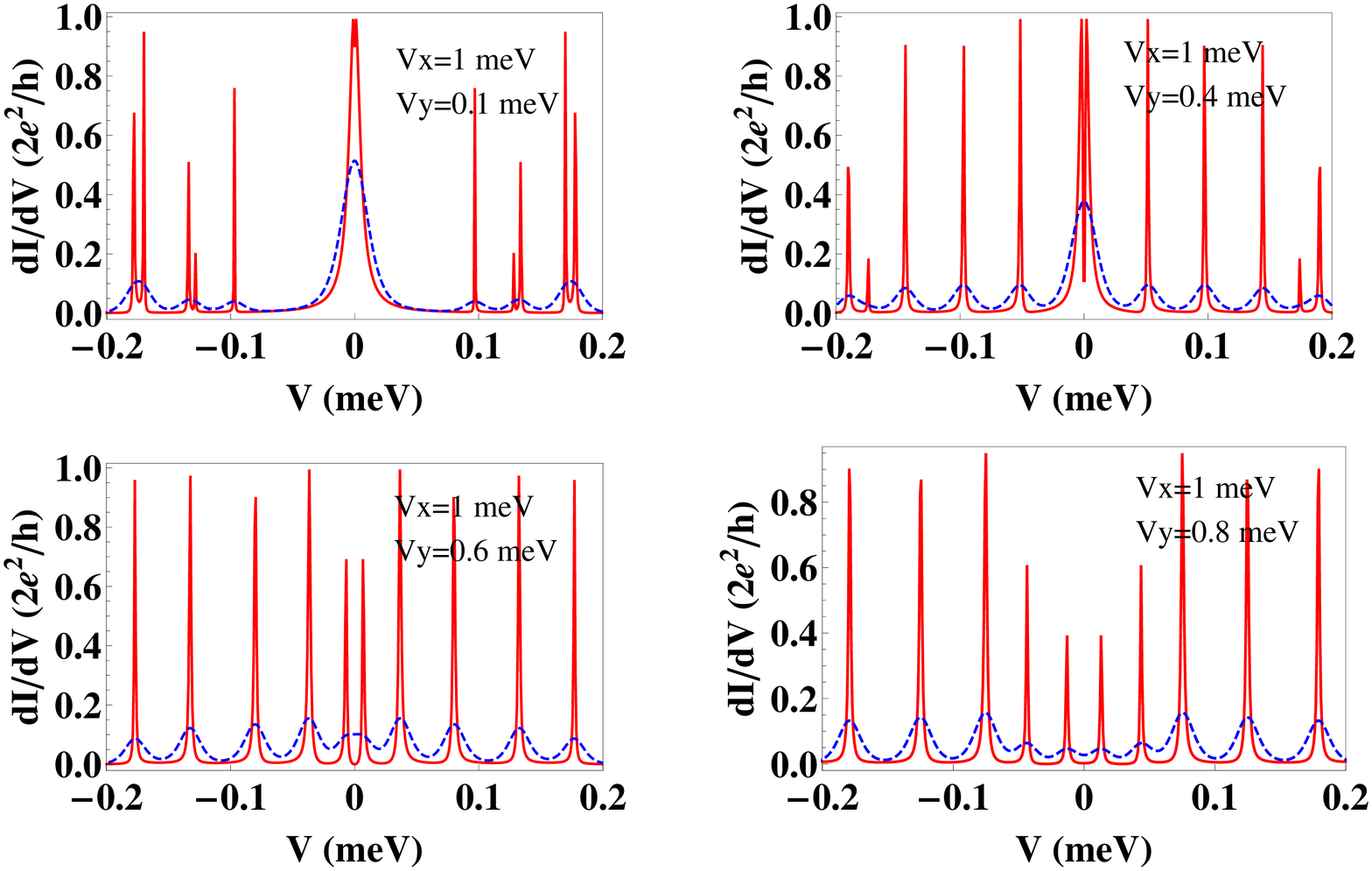}%
\caption{The differential conductance for a fixed $V_{x}=1$ meV and different
$V_{y}=0.1,0.4,0.6$ and $0.8$ meV. The red solid and blue dashed denote
temperatures $T=0$ and $60$ mK, respectively. At $V_{x}=1$ meV, the system is
in the topological phase with quantized ZBCP. $V_{y}$ will reduce the
quasi-particle gap and hence suppress the ZBCP. Finite temperature also
reduces the ZBCP. ($U=38$ meV, $L=4.5\mu m,$ spin-orbit coupling $\alpha=0.2$
meV\AA , induced pairing potential $\Delta=0.5$ meV)}%
\label{fig3}%
\end{center}
\end{figure}
%EndExpansion

%TCIMACRO{\FRAME{ftbpFU}{3.1142in}{1.094in}{0pt}{\Qcb{(Left Panel) ZBCP as a
%function of $V_{y}$ for a fixed $V_{x}=1$ meV at zero temperature denoted by
%the blue solid line. ZBCP shows a plateau for small extent of $V_{y}$ and is
%then suppressed by larger $V_{y}.$ The red dashed line denotes the
%quasiparticle energy gap. The peaks appear after gap closure. (Right Panel)
%Blue solid and red dashed lines denote different temperature $T=60$ and $120$
%mK, respectively. At finite temperature the ZBCP is suppressed. ($L=4.5$ $\mu
%m,$ spin-orbit coupling $\alpha=0.2$ meV\AA , induced pairing potential
%$\Delta=0.5$ meV)}}{\Qlb{fig4}}{fig4.eps}%
%{\special{ language "Scientific Word";  type "GRAPHIC";  display "USEDEF";
%valid_file "F";  width 3.1142in;  height 1.094in;  depth 0pt;
%original-width 15.7569in;  original-height 4.6838in;  cropleft "0";
%filename 'fig4.eps';file-properties "XNPEU";}} }%
%BeginExpansion
\begin{figure}
[ptb]
\begin{center}
\includegraphics[
height=1.094in,
width=3.1142in
]%
{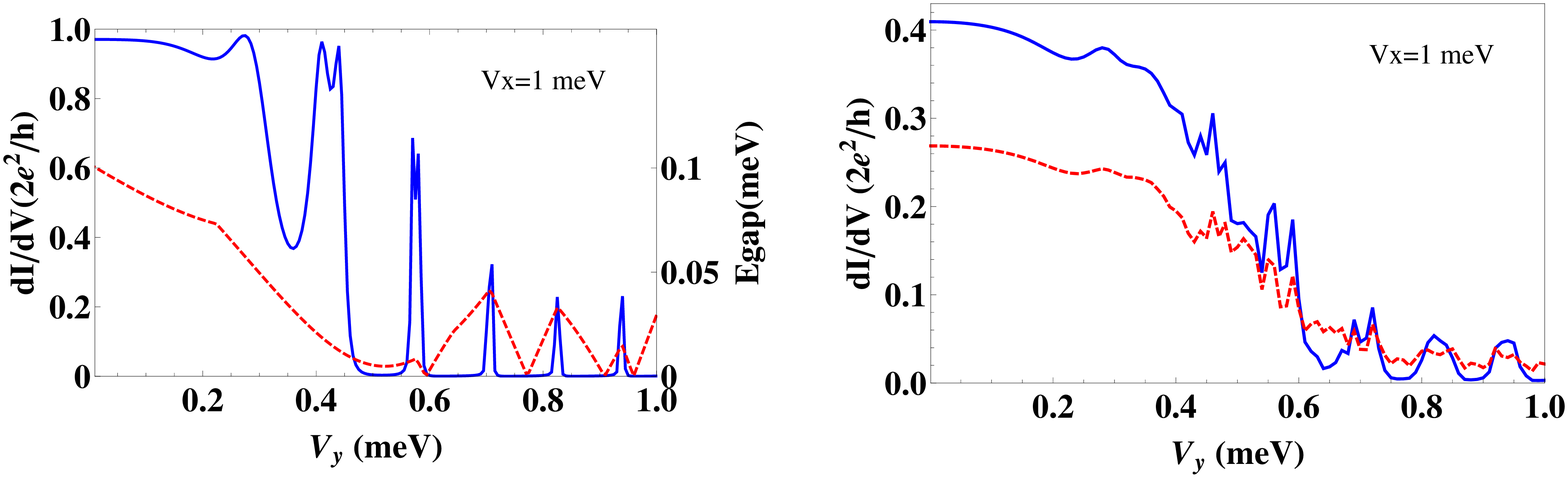}%
\caption{(Left Panel) ZBCP as a function of $V_{y}$ for a fixed $V_{x}=1$ meV
at zero temperature denoted by the blue solid line. ZBCP shows a plateau for
small extent of $V_{y}$ and is then suppressed by larger $V_{y}.$ The red
dashed line denotes the quasiparticle energy gap. The peaks appear after gap
closure. (Right Panel) Blue solid and red dashed lines denote different
temperature $T=60$ and $120$ mK, respectively. At finite temperature the ZBCP
is suppressed. ($L=4.5$ $\mu m,$ spin-orbit coupling $\alpha=0.2$ meV\AA ,
induced pairing potential $\Delta=0.5$ meV)}%
\label{fig4}%
\end{center}
\end{figure}
%EndExpansion

\begin{figure}[ptb]
\begin{center}
\includegraphics[
scale=0.6
]{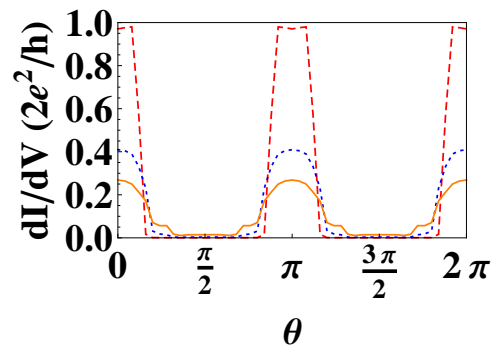}
\end{center}
\caption{ ZBCP as a function of angle $\theta=tan^{-1}(V_{y}/V_{x})$ for a
fixed $|\bm V|=1$ meV at $T=0,60,120$ mK denoted by the red dashed, blue
dotted and orange solid lines, respectively. ($L=4.5\,\mu$m, spin-orbit
coupling $\alpha=0.2$ meV\AA , induced pairing potential $\Delta=0.5$ meV).}%
\label{fig5}%
\end{figure}

Having provided reasonably realistic probable explanations for the two
observed puzzling features of Ref.[\onlinecite{1}], namely, the suppressed
values of ZBCP (Fig.\ref{fig1}) and the splitting of ZBCP (Fig.\ref{fig2}), we
now consider the effect of a Zeeman field $V_{y}$ in the spin-orbit coupling
direction (i.e. transverse to the wire length). If $V_{y}\gg V_{x},$ we expect
the ZBCP to disappear even if the system remains in the TS phase in accordance
with the invariant Pfaffian calculation \cite{17}. This is because a large
$V_{y}>\Delta$ is known to suppress the quasiparticle gap rendering any end
state completely delocalized across the wire.\cite{4} In Figs.\ref{fig3} and
\ref{fig4} we present our predicted results for the tunnel conductance for
different finite values of $\left(  V_{x},V_{y}\right)  .$ It is clear from
Fig.\ref{fig3} that increasing $V_{y}$ first suppresses the value of the ZBCP
(at finite $T=60$ mK), eventually making it disappear (as expected). For our
chosen parameters for the system, the ZBCP essentially disappears for
$V_{y}\sim\Delta\sim0.5$ meV. In Fig.\ref{fig4} we plot the actual value of
the differential conductance at the ZBCP as a function of $V_{y}$ for
$V_{x}=1$ meV for $T=60$ and $120$ mK, and it is clear that ZBCP would
disappear for $V_{y}\sim V_{x},$ particularly at higher temperatures. We
therefore predict that the experimentally observed ZBCP signature in
Ref.[\onlinecite{1}], for our estimates of the parameters of the experiment,
should essentially completely disappear for the tilt angle $\theta\gtrsim45%
%TCIMACRO{\U{b0}}%
%BeginExpansion
{{}^\circ}%
%EndExpansion
.$ An interesting notable (and experimentally verifiable) feature apparent in
Fig.4 is that the ZBCP is quite immune to a finite $V_{y}$ until $V_{y}$
becomes reasonably large ($>0.4$ meV for our chosen parameters) when it is
suppressed reasonably quickly. The recent experimental study of the
ZBCP\cite{1}, in fact, has studied the evolution of the ZBCP as a function of
transverse magnetic field $V_{y}$, by varying the angle $\theta=\tan
^{-1}(V_{y}/V_{x})$ in the plane of the wire, while holding the magnitude
constant. In Fig.\ref{fig5} we present our numerical results for the ZBCP as a
function of $\theta$ and we find oscillations of the ZBCP that are consistent
with the experimental results in Ref.[\onlinecite{1}].

\begin{figure}[ptb]
\begin{center}
\includegraphics[
scale=0.4
]{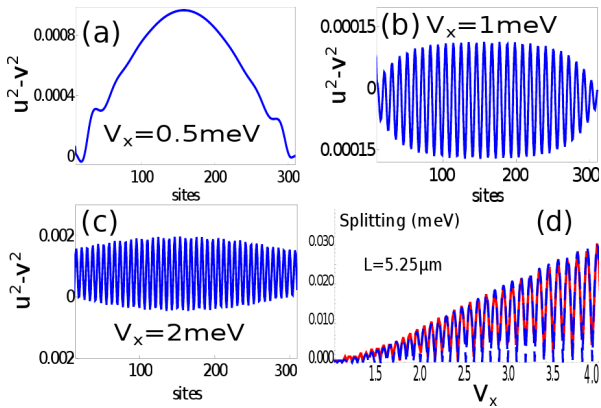}
\end{center}
\caption{ Panels (a),(b) and (c) show the evolution of the charge density
$u^{2}-v^{2}$ of the split MFs generated by overlap for $L=4.5\,\mu m$ wire
with different values of applied Zeeman-field $V_{x}$. Increasing $V_{x}$
increases the overlap between MFs, which in turn leads to a finite charge
$|u|^{2}-|v|^{2}\neq0$ associated with the split Majorana fermion. Panel (d)
shows the splitting of the Majorana fermion energy resulting from the same
overlap as a function of Zeeman-field $V_{x}$. The red line is calculated
result, while the blue dashed line is a fit to the analytically expected
overlap as discussed in the text. }%
\label{fig6}%
\end{figure}

\section{\bigskip Proposed new experiment}

The splitting of the MFs shown in Fig.\ref{fig2} depends on the applied Zeeman
field, $V_{x}$, and arises from overlap of the MF wave-functions localized at
the two wire ends. The overlap of the MF wave-functions leads to Andreev
states at non-zero energy which are no longer neutral i.e. $|u|^{2}%
-|v|^{2}\neq0$ \cite{interferometry} in contrast to the MF mode itself, which
being a precise zero-energy mode must be precisely a neutral zero-charge
quasiparticle. The total ground state charge density in the wire can in
principle be calculated using the expression
\begin{equation}
\langle{\rho(x)}\rangle=\sum_{n:E_{n}<0}\tanh{\frac{E_{n}}{2k_{B}T}}%
(|u_{n}(x)|^{2}-|v_{n}(x)|^{2}),
\end{equation}
where the Bogoliubov quasiparticle operators with energy $E_{n}$ have the form
$\Psi_{n}^{\dagger}=\int dxu_{n}(x)c^{\dagger}(x)+v_{n}(x)c(x)$. However, the
charge density $\langle{\rho(x)}\rangle$ is influenced by the presence of
disorder and one cannot in general separate the contribution of the background
charge density from the charge density resulting from the splitting of MFs.
The contribution of the split MFs to the charge density can be separated 
from the fixed background charge density (which contains the contribution from 
impurity potentials) by controlling the occupancy of the split MF modes by coupling 
the ends of the superconducting system to 
 an external normal fermionic lead. 
By changing the chemical potential of the normal fermionic lead, one changes the occupancy of the 
split MF modes, which leads to a change in the charge density in the case when the split MF modes 
are charged. The change in the charge density as a function of chemical potential of 
the normal fermion lead, which we refer to as "non-local compressibility" $\frac{\delta\langle{\rho(x)}\rangle}{\delta\mu_{lead}}$ separates out the effect of disorder from the charge of the split MFs.
To calculate the non-local compressibility, we integrate 
out the normal fermionic lead creates a self-energy for the Bogoliubov states
of the form
\begin{equation}
\Sigma_{lead}(\omega)=i\Gamma\tau_{z}\text{sgn}(\omega-\mu_{lead}),
\end{equation}
where $\Gamma$ is related to the transparency of the contact between the lead
and the superconductor. This leads to an expression for the lead chemical
potential $\mu_{lead}$ dependent charge density in the semiconductor
\begin{equation}
\langle{\rho(x)}\rangle(\mu_{lead})=\sum_{n:E_{n}>0}\tanh{\frac{E_{n}%
-\mu_{lead}}{2k_{B}T}}(|v_{n}(x)|^{2}-|u_{n}(x)|^{2}).
\end{equation}
The non-local compressibility defined as
\begin{equation}
\frac{\delta\langle{\rho(x)}\rangle}{\delta\mu_{lead}}=\frac{1}{T}%
\sum_{n:E_{n}>0}\text{sech}^{2}{\frac{E_{n}-\mu_{lead}}{2k_{B}T}}%
(|v_{n}(x)|^{2}-|u_{n}(x)|^{2}),
\end{equation}
can in principle\cite{yacoby} be measured-using single-electron-transistor
(SET) spectroscopies in lock-in-mode \cite{yoo,amir_comp}. At low temperatures
$T$, the compressibility singles out the split MF states near the chemical
potential $E_{n}\sim\mu_{lead}$. When the value of $\mu_{lead}$ coincides with
the energy $E_{n}$, there would be a peak in the non-local compressibility.
The spatial profile of this peak is related to $|u|^{2}-|v|^{2}$, which as
shown in Fig.\ref{fig6} would show an oscillatory profile in space whose
magnitude increases with the strength of the Zeeman field $V_{x}$. Here one
has ignored the screening properties of the superconducting density of states.
However, if only part of the nanowire is covered by the superconductor and the
screening length of carriers in the nanowire is long compared to the
oscillation wave-vector shown in Fig.\ref{fig6}, the electrostatic screening
effects of this charge density are not expected to be significant.
Fig.\ref{fig6}(d) shows the MF energy splitting resulting from the Zeeman
splitting. This splitting (shown by the red curve) also oscillates as a
function of the Zeeman splitting $V_{x}$. The splitting results from the
overlap of the MF wave-functions \cite{meng}, which in the case of
semiconductor nanowires has been shown to have the form\cite{4} $\psi
(L/2)=\psi_{0}e^{i(i\xi^{-1}(V_{x})+k_{F}(V_{x}))L/2}$. The resulting MF
overlap and splitting have the form splitting$~e^{-L/\xi(V_{x})}(a+bV_{x}%
)\cos{(k_{F}(V_{x})L+c+dV_{x})}$ where $a,b,c,d$ are fitting parameters. As
seen in Fig.\ref{fig6}(d), we find the parameters $(a,b,c,d)$ can be chosen to
fit the $V_{x}$ dependence over a large range. This shows that the splitting
of the low-energy eigenstates arising from the end of the nanowires shown in
Fig.\ref{fig6} is consistent with resulting from the overlap of the localized
end MFs.

The above discussion and the results presented in Fig.6 provide an exciting
possibility for a scanning SET-based measurement and detection of the MF mode
in semiconductor hybrid structures by measuring directly the charge associated
with the split ZBCP modes in large Zeeman fields (where the end MFs overlap
leading to MF-splitting), and then extrapolating this back to zero-splitting
which should correspond to the neutral MF. This is by no means an easy
experiment, but in principle, it is doable using existing SET techniques, and
the spatially-resolved MF information coming out of such an experiment would
be invaluable in understanding the details not only about the existence of the
MF, but also about its wavefunction and its spatial location.

\section{Discussion: disorder and other effects}
The theoretical work and the numerical results presented in this work are based on
a simple effective model which neglects many complications of the real system
studied in the experiment.  For example, although it is straightforward to include
disorder in our model by adding a random one-particle potential in the tight-binding
Hamiltonian, we have ignored providing results including effects of disorder because
disorder effects have already been studied elsewhere in the literature, \cite{disorder} 
and the details of the disorder operational in the
experimental systems are not known.  The main effect of disorder is to suppress the
topological gap (and to eventually destroy it for disorder strength comparable to
the gap size), and the specific experimental observations under consideration in the
current work (namely, the suppression of ZBCP height, the Majorana splitting, the
magnetic field effects, etc.) are unlikely to be qualitatively modified by disorder, 
particularly for the relatively clean wires studied in Ref.[\onlinecite{1}].
 Our work also neglects any orbital coupling introduced by the magnetic field which
could affect some of the experimental findings at very large fields where the
magnetic length could be shorter than the wire width.  Another limitation of the
current theory is to approximate the tunneling measurement simply by a single
tunneling barrier height U (and the finite wire length) without considering the
actual details of the tunnel junction used in the experiments.  This would
necessitate a careful modeling of the tunnel contact which is well beyond the scope
of the current minimal model.  A full understanding of the experimental details must
await a more quantitative realistic modeling of the actual
semiconductor-superconductor hybrid structure used in the laboratory, which may,
however, be challenging since many details (e.g. the nature of various interfaces)
are unknown.

Two specific new results arising from our theoretical work are partially in
agreement with observations: the large suppression of the ZBCP height compared with
the pristine theoretical quantization ($2e^2/h$) and the splitting of the ZBCP at
large magnetic fields.  Even for these two predictions, questions could be raised if
alternate mechanisms are operational.  The strong dependence of the actual ZBCP on
the wire length and the tunnel barrier that we find in our results establishes that
temperature by itself is insufficient to explain the quantitative height of the
ZBCP, but neither the precise effective wire length nor the tunnel barrier height
are known, again making a comparison between theory and experiment problematic at a
quantitative level.  False color images can easily be produced by adjusting various
parameters making theory and experiment look alike, but this may not be particularly
useful in view of the large number of unknowns in the problem.  The Majorana
splitting at high magnetic fields is inevitable due to the increasing magnitude of
the dimensionless separation between the two end Majorana modes (as a ratio of the
coherence length which increases monotonically with increasing field due to the
suppression of the gap), and thus all experimental observations of the Majorana in
real systems with finite wire lengths must eventually reflect this splitting.  The
Majorana splitting of the ZBCP should not behave as a simple linear function of the
magnetic field distinguishing it clearly from any trivial Zeeman splitting.  A clear
prediction of the theory is that the splitting should oscillate in some manner, and
such oscillations have not yet been seen experimentally.  Whether the
non-observation of the Majorana splitting oscillation is a genuine feature or is due
to disorder and thermal broadening is unclear at this stage.  Only further work can
clarify this question.

\section{Conclusion}

Before concluding, we point out that there are various resonance structures in
our numerical results which arise from the sharpness of our confinement and
transport models, which are completely non-universal and non-topological in
nature. Such resonant structures in the current-voltage characteristics are
well-known in 1D systems\cite{4,15} and arise from various resonances in the
transmission and reflection coefficients. Presence of discrete impurities may
lead to additional non-topological resonant structures. These resonant
structures will shift around with gate voltage and magnetic field with the
ZBCP being the only universal topologically robust feature in the data.

We mention that although the results presented in this work are restricted to
the one-subband strict 1D limit (i.e. very large inter-subband gap energy)
with no disorder, we have carried out some representative calculations for
multisubband-occupied disordered wires finding qualitatively similar results,
leading to our belief that our results and conclusions presented in this work
continue to apply qualitatively in more realistic multisubband wires in the
presence of finite disorder (provided a TS phase can be realized in the
system). While any detailed quantitative comparison between experiment and
theory must await more realistic modelling of the actual SM/SC systems
utilized in Ref.[\onlinecite{1}], our current work decisively demonstrates
that the suppression of the ZBCP well below the canonically quantized value,
splitting of the ZBCP at high Zeeman fields, and the suppression of the ZBCP
in the presence of a Zeeman field along the spin-orbit direction are all
expected theoretical features of the SM/SC Majorana system proposed in
Refs.[\onlinecite{2,3,4,5}] and observed in Ref.[\onlinecite{1}]. In addition,
we propose a possible new experiment using the scanning SET spectroscopy,
which, in principle, can provide information about the spatial location of the
MF and its wavefunction by measuring the effective charge on the Andreev bound
states associated with the splitting of the MFs in a finite wire at higher
values of the Zeeman field.

Our work here is an extension and generalization of the standard theory for
the Majorana mode \cite{2,3,4,5} to finite wire lengths at finite
temperatures. This is important for the comparison with the experiment
\cite{1} at a qualitative level since this directly tells us the extent to
which the standard zeroth order theory captures the essential features of the
experiment. We do not introduce any new ingredient except to extend the
standard theory to the experimental finite systems at finite temperatures so
that an honest and direct assessment is possible about the key issue of
whether the experiment of Ref.[1] really explores the theoretical predictions
of Refs.2-5. Such a comparison with the extended standard theory is essential
to establish future directions of research in this subject. We do find several
surprises in our results of applying the standard theory to finite systems
corresponding to finite length nanowires used in the experiments. For example,
the ZBCP never reaches its expected quantized value in finite length wires
even at very low temperatures because the details of the tunneling barrier now
become an important variable. Another somewhat unexpected result, which should
be checked in future experiments, is an oscillatory splitting of the MF as a
function of the effective Zeeman splitting at high values of the magnetic
field where the topological gap is sufficiently suppressed so that the
coherence length is larger than the wire length.

We conclude by stating that we have established in this work that the two
puzzling features of the likely experimental observation of the Majorana modes
in 1D InSb nanowire\cite{1} following earlier theoretical
proposals\cite{2,3,4,5} can be explained by including finite wire length,
finite temperature, and finite tunneling barrier effects in the theory. We
have also made specific predictions on how the zero-bias-conductance peak will
be suppressed in the presence of a finite magnetic field in the transverse
direction and how the Majorana-induced zero-bias-conductance-peak will split
at high Zeeman splitting in a finite wire due to the overlap between the
Majorana modes at the two ends of the wire. Our work does not prove that the
experiment reported in Ref.[1] is indeed the experimental discovery of the
long-awaited Majorana fermion, particularly since unknown effects could, in
principle, conspire to give a weak zero bias peak which follows for unknown
reasons the phenomenology observed in the experiment and theoretically modeled
in the current work. What our work does provide is compelling support for the
claim that the observation of Ref.[1] is indeed consistent with the
theoretically predicted existence of the MF in semiconductor hybrid
structures, and the apparent anomalies in the data of Ref.[1] (suppressed
ZBCP, high-field splitting of the ZBCP, etc.) are all completely in accord
with the zeroth order theory. The absolute confirmation of the solid state MF
discovery must await a direct demonstration of the non-Abelian nature of these
quasiparticles through an interferometry measurement which however is unlikely
to be either easy or quick. Meanwhile, further observations of ZBCP in other
nanowires by different groups and various consistency checks between theory
and experiment, as carried out in the current work, would go a long way in
establishing that perhaps the elusive Majorana has finally shown up in a most
unusual place, in a semiconductor placed on a superconductor in the presence
of Zeeman spin splitting.

This work is supported by Microsoft Q and JQI-NSF-PFC. J. S. acknowledges
support from the Harvard Quantum Optics Center.

\textit{Note added}: After the posting of our work, the results of
Ref.[\onlinecite{1}] appeared on line in its published form\cite{leo_science}.
All our results, discussion, and conclusion with respect to
Ref.[\onlinecite{1}] apply equally well to the published results in
Ref.[\onlinecite{leo_science}]. Thus, the case in favor of the likely
experimental observation\cite{1,leo_science} of the possible signatures for
the existence of the proposed\cite{2,3,4,5} Majorana modes in SM/SC hybrid
structures is further enhanced by our theoretical results. We point out,
however, that at best the observations of Refs.[\onlinecite{1,leo_science}]
establish only the necessary conditions for the existence of the long-sought
emergent Majorana modes in solid state systems. Much more work would be
needed, including the observation of similar effects in other semiconductor
nanowires with strong spin-orbit coupling (e.g. InAs) and the experimental
demonstration of the sufficient conditions for the existence of the Majorana
modes involving the observation of the fractional Josephson
effect\cite{3,10,kwon} and/or the non-Abelian braiding\cite{braiding}, before
one can compellingly claim to have discovered the elusive Majorana
quasiparticles in solid state systems.

\bigskip\


\begin{thebibliography}{99}                                                                                               %

\bibitem {1}V. Mourik,  K. Zuo,  S. M. Frolov,   S. R. Plissard,   E. P. A. M. Bakkers,
and L. P. Kouwenhoven Science, 1222360 (2012).

%\bibitem {1}L.P. Kouwenhoven, talk D44.00003 presented at the APS March
%Meeting in Boston (2012).


\bibitem {2}J. D. Sau, R. M. Lutchyn, S. Tewari, S. Das Sarma, Phys. Rev.
Lett. \textbf{104}, 040502 (2010).

\bibitem {4}J. D. Sau, S. Tewari, R. Lutchyn, T. Stanescu and S. Das Sarma,
Phys. Rev. B \textbf{82}, 214509 (2010).

\bibitem {3}R. M. Lutchyn, Jay D. Sau, S. Das Sarma, Phys. Rev. Lett.
\textbf{105}, 077001 (2010) .


\bibitem {5}Y. Oreg, G. Refael, F. V. Oppen, Phys. Rev. Lett. \textbf{105},
177002 (2010).

\bibitem {R1}E. Reich, Nature \textbf{483}, 132 (2012); F. Wilczek, Nature
\textbf{486}, 195 (2012); P. Brouwer, Science \textbf{25,} 989 (2012); R.
Service, Science\textbf{ 332}, 193 (2012); R. Wilson, Physics Today
\textbf{65}, 14 (2012)

\bibitem {6}R. M. Lutchyn, T. Stanescu, S. Das Sarma, Phys.Rev.Lett.
\textbf{106}, 127001 (2011).


\bibitem {10}A. Y. Kitaev, Physics-Uspekhi \textbf{44}, 131 (2001).


\bibitem {9}K. Sengupta, I. Zutic, H. Kwon, V. M. Yakovenko, and S. Das Sarma,
Phys. Rev. B \textbf{63}, 144531 (2001).




\bibitem {7}T. Stanescu, R. M. Lutchyn, S. Das Sarma, Phys. Rev. B
\textbf{84}, 144522 (2011)

\bibitem {8}K. Flensberg, Phys. Rev. B. 82, 180516(R) (2010)


\bibitem {11}Jay D. Sau, Sumanta Tewari, and S. Das Sarma, Phys. Rev. B
\textbf{85}, 064512 (2012)

\bibitem {12}K. T. Law, Patrick A. Lee, and T. K. Ng, Phys. Rev. Lett.
\textbf{103}, 237001 (2009).

\bibitem {ramon}E. Prada, P. San-Jose, R. Aguado, arXiv:1203.4488 (2012); J.
Lim, L. Serra, R. Lopez, R. Aguado, arXiv:1202.5057 (2012)

\bibitem {wimmer}M. Wimmer, A.R. Akhmerov, J.P. Dahlhaus, C.W.J. Beenakker,
New J. Phys. \textbf{13}, 053016 (2011).

\bibitem {13}S. Nadj-Perge, V. S. Pribiag, J. W. G. van den Berg, K. Zuo, S.
R. Plissard, E. P. A. M. Bakkers, S. M. Frolov, L. P. Kouwenhoven. e-print arXiv:1201.3707

\bibitem {bolech_demler}C. J. Bolech, E. Demler, Phys. Rev. Lett. \textbf{98},
237002 (2007).

\bibitem {meng}M. Cheng, R. M. Lutchyn, V. Galitski, S. Das Sarma, Phys. Rev.
Lett. \textbf{103}, 107001 (2009).

\bibitem {nilsson}H. A. Nilsson, P. Samuelsson, P. Caroff, and H. Q. Xu Nano
Letters \textbf{12}, 228 (2012).

\bibitem {rokhinsonprivate}L. P. Rokhinson, private communication .

\bibitem {marcusprivate}C. M. Marcus, private communication.

\bibitem {doh}Y. J. Doh et al., Science \textbf{309}, 272 (2005).

\bibitem {16}G. E. Blonder, M. Tinkham, and T. M. Klapwijk, Phys. Rev. B
\textbf{25}, 4515 (1982).

\bibitem {14}Meng Cheng, Roman M. Lutchyn, Victor Galitski, S. Das Sarma,
Phys. Rev. B \textbf{82}, 094504 (2010).

\bibitem {17}P. Ghosh, J. D. Sau, S. Tewari, S. Das Sarma, Phys. Rev. B,
\textbf{82}, 184525 (2010).

\bibitem {interferometry}J. D. Sau, S. Tewari, S. Das Sarma, Phys. Rev. B
\textbf{84}, 085109 (2011)

\bibitem {yacoby}A. Yacoby, private communication.

\bibitem {yoo}M. J. Yoo et al., Science \textbf{276}, 579 (1997).

\bibitem {amir_comp}S. Ilani, A. Yacoby, D. Mahalu, H. Shtrikman, Science
\textbf{292}, 1354 (2001); V. Venkatachalam, A. Yacoby, L. Pfeiffer, K. West,
Nature \textbf{469}, 185 (2011)

\bibitem{disorder}A. C. Potter, P. A. Lee, Phys. Rev. B 83, 184520 (2011);
T. D. Stanescu, R. M. Lutchyn, S. Das Sarma, Phys. Rev.
B 84, 144522 (2011); R. M. Lutchyn, T. D. Stanescu, S.
Das Sarma, Phys.Rev.Lett. 106, 127001 (2011); R. M.
Lutchyn, T. D. Stanescu, S. Das Sarma, Phys. Rev. B 85, 140513(R) (2012); P. W. Brouwer, M. Duckheim, A. Romito, F.
von Oppen , Phys. Rev. Lett. 107, 196804 (2011); P. W.
Brouwer, M. Duckheim, A. Romito, F. von Oppen ,Phys.
Rev. B 84, 144526 (2011); J. D. Sau, S. Tewari, S. Das
Sarma, Phys. Rev. B \textbf{85}, 064512 (2012).

\bibitem {15}Song He and S. Das Sarma, Phys. Rev. B \textbf{48}, 4629 (1993).

\bibitem {leo_science}V. Mourik, et al., arXiv:1204.2792; Science Express
April 12, on line 1222360 (2012).

\bibitem {kwon}H.J. Kwon et al., Eur. Phys. J. B \textbf{37}, 349 (2004)



\bibitem {braiding}F. Hassler, A. R. Akhmerov, C.-Y. Hou, C. W. J. Beenakker,
New J. Phys. \textbf{12}, 125002 (2010); J. D. Sau, S. Tewari, S. Das Sarma,
Phys. Rev. A \textbf{82}, 052322 (2010); J. Alicea, Y. Oreg, G. Refael, F. von
Oppen, M. P. A. Fisher, Nature Physics \textbf{7}, 412 (2011); J. D. Sau, D.
J. Clarke, S. Tewari Phys. Rev. B \textbf{84}, 094505 (2011); B. I. Halperin,
Y. Oreg, A. Stern, G. Refael, J. Alicea, and F. von Oppen, Phys. Rev. B
\textbf{85}, 144501 (2012); B. van Heck, A.R. Akhmerov, F. Hassler,
M.Burrello, C.W.J. Beenakker, New J. Phys. \textbf{14} 035019 (2012).
\end{thebibliography}
\end{document}